\documentstyle{article}

\title{{\large{\bf Calculations of the Local Density of States for Some Simple Systems}}}

\bigskip

\author{  {\bf Alexandre Losev}
\\{\it Bulgarian Academy of Sciences, Institute of General and Inorganic Chemistry,}
\\{\it Sofia 1113, Bulgaria}
\\and 
\\{\bf Stoyan Vlaev}
\\{\it Escuela de Fisisca,  Universidad Autonoma de Zacatecas, 
98068 Zacatecas, Mexico}}

\begin{document}
\maketitle

\bigskip
\bigskip

\section*{\rm{Abstract}}
 A recently proposed convolution technique for the calculation of local densities of states (LDOSs) is described more thoroughly and new results of its application are presented. For separable systems the exposed method allows to construct the LDOS for a higher dimensionality out of lower dimensional parts. Some practical and theoretical aspects of this approach are also discussed.

PACS 71.20.-b, 71.15.-m, 73.20.-r

\section{\bf{Introduction}}

\bigskip

	Recently we proposed and demonstrated a new convolution technique for the calculation of the local density of states [1,2]. The idea of this novel approach is intuitively accessible, so we had not yet exposed its mathematical description. The analytic formulation offers the advantage of a more comprehensive view about the interplay of parameters in the utilized models, which could be beneficial with  view to further developments. Exact analytic results being rather scarce  in this area [3-6], the calculations of the LDOS and other characteristics rely on numerical schemes but usually they become a cumbersome task even for some of the most simple systems. In consequence, a wealth of alternatives has been produced and among them the ones exploiting  tight-binding (TB) models appear as particularily useful [7-9]. The TB approximation also provides a convenient framework for the present work. The approach considered here proposes to utilize in a constructive manner the already know LDOSs in analytic or numeric form, no matter how they have been arrived at.

\section{\bf{Model and Method}}

Within the tight-binding model the Hamiltonian of a simple D-dimensional crystal is written as $H= \alpha + 2 \sum_{i} \beta_{i} cos(k_{x_{i}}d_{i})$, 
$i=1..D$. Considered from a statistician's point of view, the distribution of the values generated by a form such as $H$ is a 'composition' or 'convolution' [10] of the distributions of its  additive parts which correspond to the independent variables. Thus, the eigenvalues of the system are indeed obtained by  a combination of eigenvalues corresponding to its components. Accordingly, separating the system into a one-dimensional part and a D-1 dimensional subsystem, the LDOS of a the whole system can be obtained from the subsystems with smaller dimensions through the recurrence:

$\rho^{\rm(D)}=\rho^{\rm (D-1)} * \rho^{\rm (1)}$ 

where  $\rho^{\rm (0)}=\delta$ and the sign $*$ denotes the convolution operation. The idea of this approach is presented graphically in Fig.1. For a better view  all the curves in this plot have been scaled to  a same maximum height, instead of keeping the same (unit) area. If the singularities of the one-dimensional curve  are well pronounced, they are reproduced in subsequent calculations, as it can be seen on this figure.

In order to have a more substantial derivation we proceed in the opposite direction, starting with $D=1$. In this case the Hamiltonian is $H= \alpha + 2 \beta cos(k_{x})$.The distance $d$ is assumed to be $d=1$ and the energy $\epsilon$ is expressed everywhere below through the dimensionless variable $E=(\epsilon-\alpha)/2 \beta$.

From the meaning of LDOS one can write directly $dk \sim \rho(E).dE $ i.e. $\rho \sim dk/dE$. For oscillators, as it has been noted  $ \rho \sim 1/\vert v \vert, v=dE/dk$ being the group velocity or more generally [11] $v= grad_{k}(E)$. So for this Hamiltonian $k_x=arcos(E)$ and thus

$\rho^{\rm(1)}={1 \over \pi \sqrt{(1-E^2)}}$,

the coefficient $1/\pi$ being written in order to scale the overall density to a unit, assuming it is equal to zero outside the energy interval (-1,1).
In two dimensions $H= \alpha +2\beta (cos(k_x)+cos(k_y))$ and in the same way we obtain for the partial LDOS 

$\rho_y={1 \over \sqrt{1-(E-cos(k_y))^2}}$.

Integrating for all values of $k_y$ and substituting $k_y$ with $arcos(u)$, after inverting the integration limits, the LDOS is

$\rho^{\rm(2)}={1 \over  \pi^{2}} \int_{\tiny{-1}}^{\tiny{1}} {1 \over \sqrt{1-(E-u)^2} \sqrt{1-u^2}}du$.

The  obtained formula is the canonical expression for a convolution,

$f(x)=\int_{-\infty}^{\infty} g(x-\xi) h(\xi) d\xi$,

for which the usual notation $f=g*h$ has been adopted. In our particular case $g$ and $h$ are indeed the same function and thus we have a self-convolution, which can be of special interest), but this does not need to  be so  as our examples in the next section demonstrate.
The same consideration is applicable to higher dimensionalities as the  multiple integrations are consecutive convolutions e.g.

$ \int{\int{{1 \over \sqrt{(1-((E-u)-v))^2} \sqrt{(1-u^2)} \sqrt{(1-v^2)}}}} du dv$

would be more generally

$f(x)=\int{\int{ g((x-\xi)-\eta) h(\eta) l(\xi)}} d\eta d\xi$,

or $f=g*h*l$ 
for some functions $f,g,h$ and $l$ representing LDOSs. The advantages of this approach appear to be fairly obvious: the results for more simple systems are utilized to obtain knowledge about more complicated ones. Most often such results are available in numeric form and then  the convolution is implemented simply as the sum $f_{k}=\sum_{i} g_{k-i} h_{i}$. However a more sophisticated procedure might be needed in order to evaluate precisely the integral which represents it. In our particular case it may be noted that the quadratures for expressions containing $1/ \sqrt{(1-x^2)}$  can be effectively  simplified [12]. Also, as it is known,  a convolution can be expressed in Fourier space by a simple multiplication of the transforms and here we have $\rho^{(D)}=FT(\vert J_o^{D} \vert)$. Any way, the computational advantages of this method can be considerable as it was already pointed [1].

\section{\bf{ Results and Discussion}}

As examples of application  we shall consider briefly two cases of  semi-infinite structures and also offer a comment on the effect of an electric field in the present context.

For the simple model consisting of two infinite wires in interaction or, which is the same, an infinite chain of dimers, the convolution method readily offers a solution. The LDOS for a dimer is given [13] by the formula:

$\rho={-1 \over \pi} Imag({g \over (1-g^2 t^2)})$, 

where $g(E)=1/(E +i\gamma)$ is the Green function for an isolated atom  and $t$ is the interaction between atoms (in the dimer). The explicit expression for the LDOS is a sum of two Lorentzians centered at $\pm t$ and whose width $\gamma$ is vanishingly small. In the limit, as they tend to delta functions, the convolution produces a sum of two one-dimensional LDOSs centered now at $\pm t$:

$\rho(E) ={1 \over 2} ( \rho^{(1)}(E-t)+ \rho^{(1)}(E+t) )$.

Fig.2 presents the result for this most simple case. 
Modifying the value of the interaction $t$ would just change the distance between the delta peaks and this will be reflected in the overlap of the one-dimensional LDOS, whose width is governed by $\beta$.  An algebraic treatment for a finite variant of this  structure is found in Ref.6.  More complicated cases are accessible when the poles of the Green function are known.
Indeed, if one chooses to consider the LDOSs as consisting of Lorentzian (or Gaussian) peaks whose centers are distributed following some law, e.g. a cosine for one dimension, then the convolution would affect only the distribution, as it produces out of two such peaks a new one with a larger full width at half max.

Another case of a semi-infinite structure is the model of a slab which is infinite in along two axes and restricted to  only one side of the third. In this case the LDOS corresponding to an infinite plane should be convoluted with the LDOSs for atoms in a semi-infinite chain, which are different according to their distance from the origin [13]. This has been illustrated in Fig.3 which in fact reproduces  results presented by Haydock and Kelly and who in their turn were assessing the possibilities of their own approach by repeating a plot originally shown in Kalkstein[14]. The last curve in the plot is  essentially a replication of the lower row of curves in Fig.1 as for atoms far from the beginning of the chain the LDOS is the same as in an infinite structure. However one might presume that here we have gained some insight into the constitution of these well known results.

Adopting this view, we could reconsider the effect of a constant electric field on the LDOS [15]. Here the analytic result is known for the one-dimensional case [5] and it has to be convoluted with its an appropriate counterpart in order to obtain the  results of interest [1]. As it was pointed, a convolution can be expressed in Fourier space by a  multiplication, and this form makes it  easier to account for the effect of an electric field:  the Fourier Transform of  $J_o(2sin(\omega)/F)$, where F is the field strength along the axis, matches [16] the result for an infinite chain obtained by Davison et al.  The FT of the LDOS, being represented by a product of such factors corresponding each to an axis, for higher field strength along any one of them,  turns into  a constant $(J_o(0)=1)$, so its original is now a delta function and the dimensionality of the system is lowered. The physical aspect of this phenomenon has been considered in the work already mentioned as Ref.15.

The limitations for this way of proceeding are fairly evident as it relies on the additivness of the Hamiltonians. However some insight in the production of the LDOS has been gained. For instance, a $\rho^{(2)}$ curve now can be seen as (the passage to the limit for) a weighted sum of 'U' shapes centered at different energies. If second neighbours are to be included in the tight-binding Hamiltonian, it ceases to be additively separable but it still can be factorized, now as a product, and then the summation would include not just weighted and displaced one-dimensional 'U' shapes but ones with a different spread. This suggests that our constructive approach could perhaps offer further possibilities for the calculation of more complicated cases.

\section*{\bf{ Conclusion}}
The proposed technique allows to utilize effectively already obtained results and to  extend them further. It is able to generate new ones and  in many instances  otherwise difficult to obtain results became easily accessible. However the method relies on the separability of the Hamiltonian,  which severely limits its scope. The analytic formulation of the method suggests a novel view even if the complexity of the calculations most often defies their completion. The convolutional form has a rather intuitive meaning, which might be more intelligible than the abstract elliptic integrals expressing the same result.

\section*{\bf{ References}}
1 A. Losev, S. Vlaev and T Mishonov, J. Phys. C {\bf11} 7501 (1999)

2 A. Losev, S. Vlaev and T Mishonov, Phys stat. solidi(b)  {\bf 220} 747 (2000).

3 M. Saitoh, J. Phys. C {\bf 5 }  914 (1972).

4 P. Kirkman and J. Pendry, J. Phys. C {\bf 17}  4327 (1984).

5 S. Davison, R. English, Z. Miskovic, F. Goodman, A. Atmos and B. Burrows,  J. Phys. C {\bf 9 } 6371 (1997).

6 R. Farchioni, G. Grosso and P. Vignolo,  Phys. Rev. B {\bf{59}} 16065 (1999).

7 C. Goringe, D. Bowler and  E. Hernandez, Rep.Prog. Phys.  {\bf{60}} 1447 (1997).

8 P. Ordejon, Computational Material Sci.  {\bf{12}} 157 (1998).

9 G. Grosso and G. Parravicini, Adv. in Chem. Phys.  {\bf 62} 133(1985).

10 H. Cramer, 1946 {\it Mathematical Methods of Statistics}, Princeton, ch. 15.12.

11 C. Kittel, {\it Introduction to Solid State Physics},(John Wiley, Toronto, New York and London, 1978), formula 6.33a.

12 {\it Handbook of Mathematical Functions}, eds. M. Abramowitz and I. Stegun, (National Bureau of Standards, Wasington, 1964), formula 25.4.38.

13 S. Vlaev, Comm. of the Department of Chem. Bulgarian Acad. of Sci., vol {\bf 17} 497 (1984).

14 R. Haydock and J. Kelly, Surf. Sci. {\bf 38} 139 (1973); D. Kalkstein and P. Soven, Surf. Sci. {\bf 26} 85 (1971); seen also in R. Messmer, Surf. Sci. {\bf 106} 225 (1981).

15 A. Losev and S. Vlaev,  Phys. stat. solidi(b) {\bf 223} 627 (2001).

16 F. Oberhettinger, {\it Fourier Transforms of Distributions and their Inverses}, (Academic Press, New York and London, 1973), Part I, table 1.10, formula 390.

\section*{\bf{ Figure Captions}}

Figure 1. The LDOS of a 3D simple cubic crystal obtained by self convolution. Upper panel: the convolution of two 1D LDOS produces a 2D LDOS; Lower panel: the convolution of the 2D LDOS and 1D LDOS produces the 3D LDOS.

Figure 2. LDOS for an infinite chain of dimers  obtained by  convolution. 

Figure 3. LDOS in  a semi-infinite slab for the first three atomic layers and bulk $(z=0,1,2..\infty)$. The dotted curves present  on a smaller scale the LDOS in a semi-infinite chain while the inset shows  the 2D LDOS in a plane used to produce the curves.

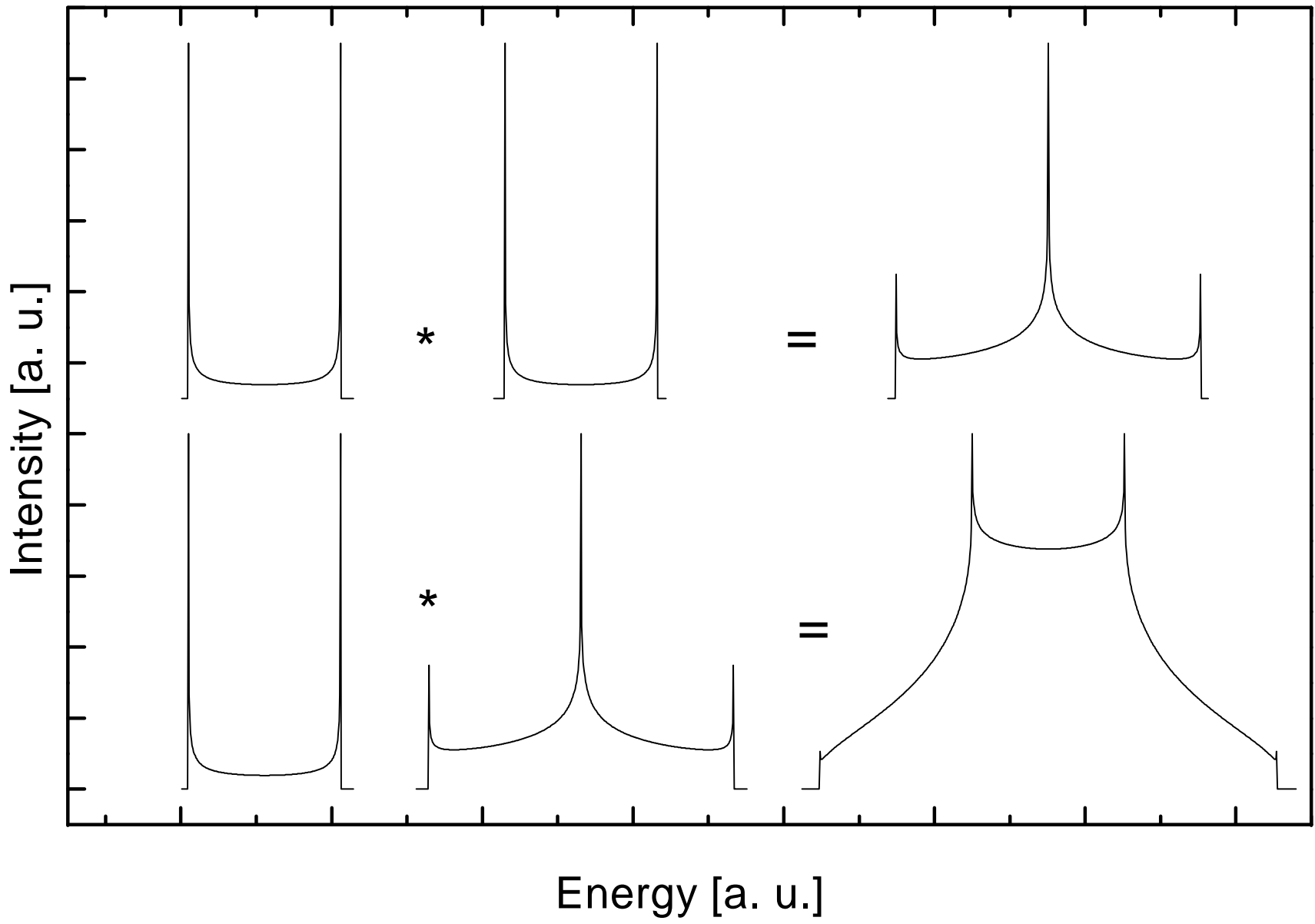
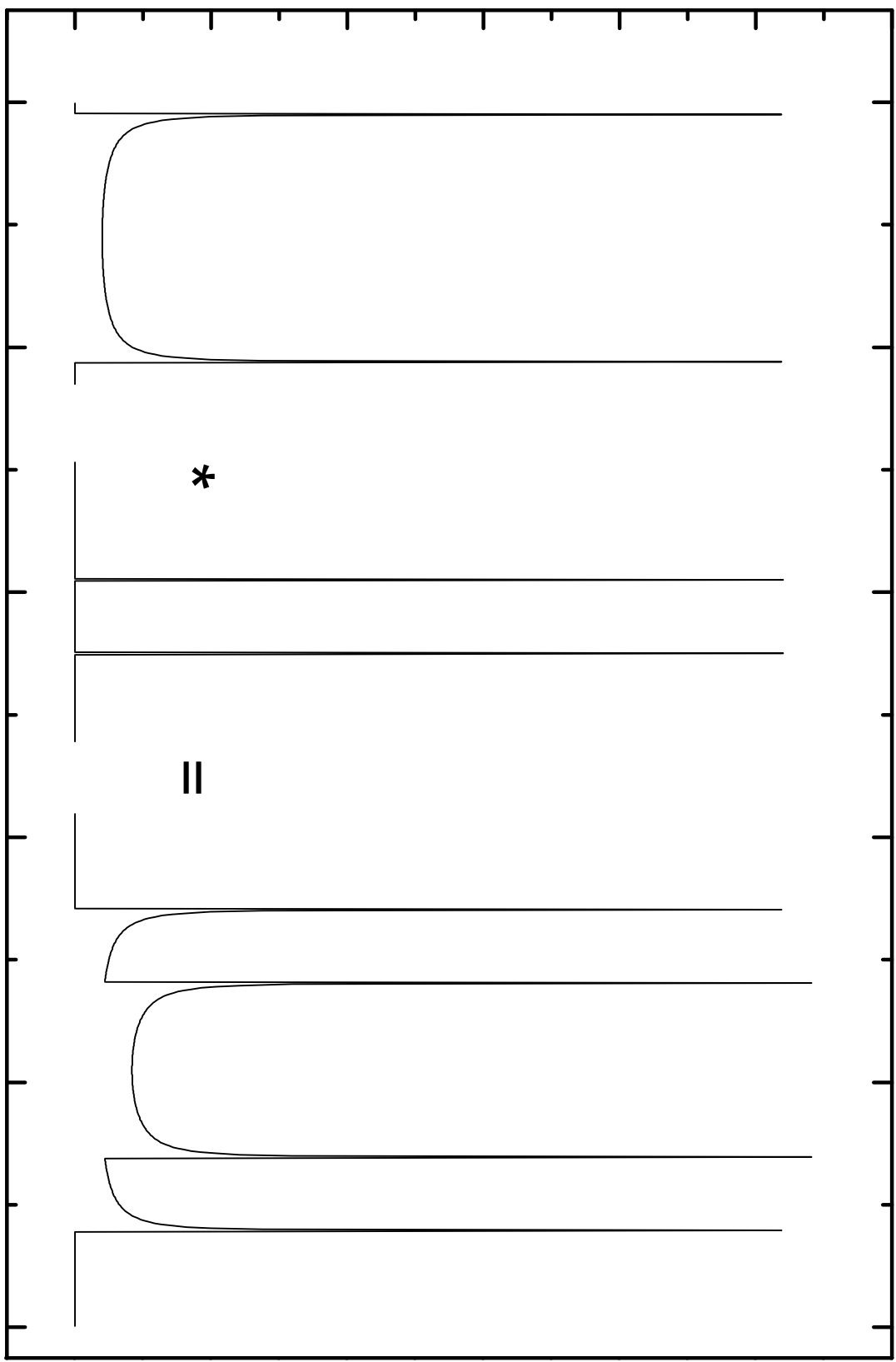
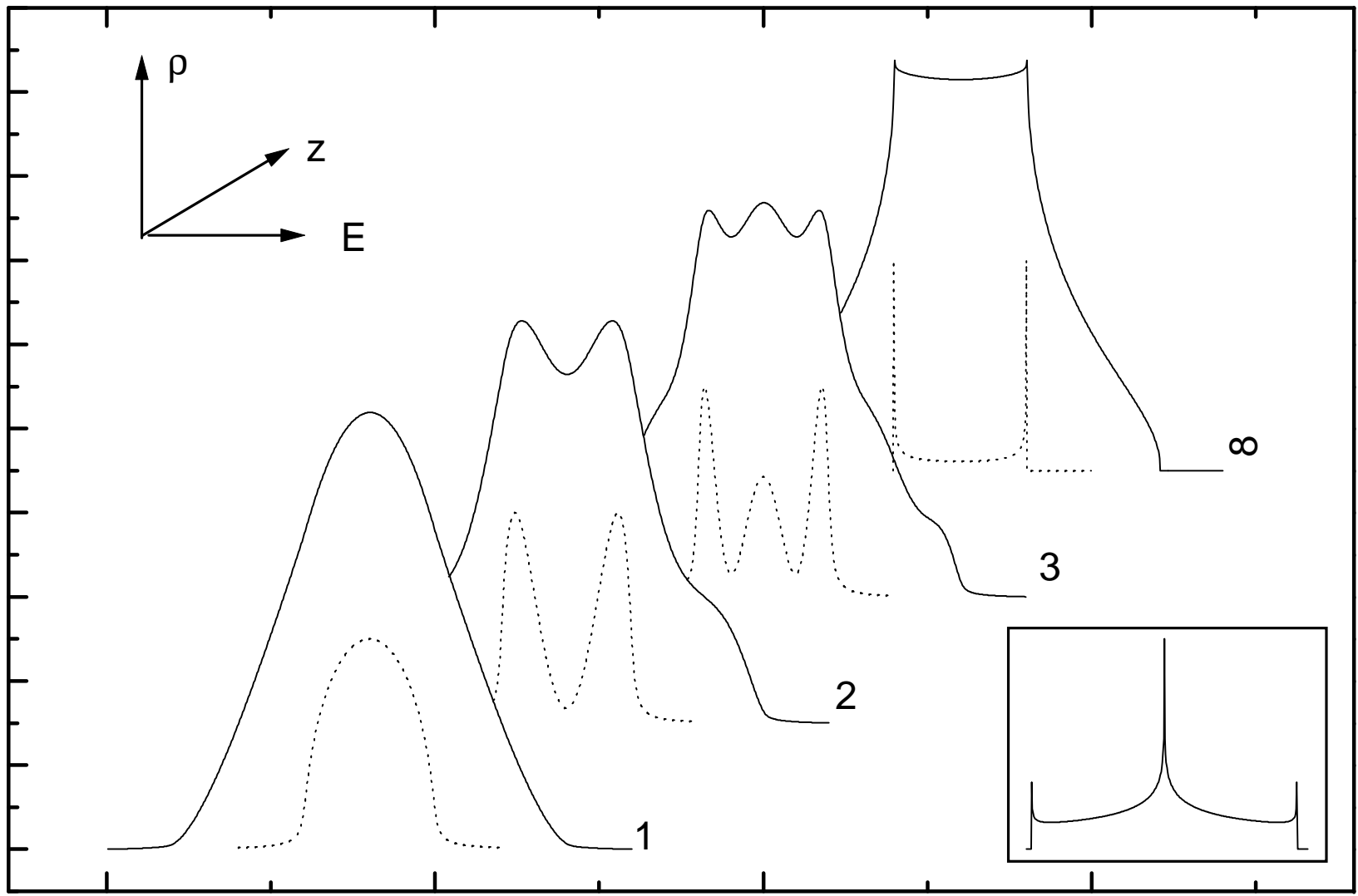
\end{document}